\documentclass[aps,twocolumn,showpacs,preprintnumbers,amsmath,amssymb,nofootinbib]{revtex4}
\usepackage{amssymb}
\usepackage{epsfig}
\usepackage{graphicx}
\usepackage{graphicx}
\usepackage{dcolumn}
\usepackage{bm}

\begin{document}
\title{Neutrino Production via $e^-e^+$ Collision at $Z$-boson Peak}
\author{Chao-Hsi Chang$^{a,b,c}$\footnote{email:zhangzx@itp.ac.cn}, Tai-Fu Feng$^{b,c,d}$\footnote{email:fengtf@hbu.edu.cn}, Feng-Yao Hou$^{b}$, Chang Liu$^{e}$
and Zan Pan$^{b}$}
\affiliation{$^a$ CCAST (World Laboratory),P.O.Box 8730, Beijing 100190, China\\
$^b$ State Key Laboratory of Theoretical Physics (KLTP),\\
Institute of theoretical
Physics, Chinese Academy of Sciences, Beijing, 100190, China\\
$^c$ Department of Physics, Hebei University, Baoding, 071002, China\\
$^d$ Department of Physics, Dalian University of Technology, Dalian,
116024, China\\
$^{e}$Department of Modern Physics, Univ. of Science and technology of China, Hefei, 230026,
China}

\begin{abstract}
The production of the three normal neutrinos
via $e^-e+$ collision at $Z$-boson peak (neutrino production
in a Z-factory) is investigated thoroughly. The differences of
$\nu_e$-pair production from $\nu_\mu$-pair and
$\nu_\tau$-pair production are presented in various aspects.
Namely the total cross sections, relevant differential cross sections and the
forward-backward asymmetry etc for these neutrinos are presented in terms of
figures as well as numerical tables.
The restriction on the room for the mixing of the three species
of light neutrinos with possible externals (heavy neutral leptons and/or stereos)
from refined measurements of the invisible width of $Z$-boson is discussed.
\end{abstract}


\maketitle

\section{Introduction}
An electron-positron collider, running at the peak of $Z$-boson
i.e. at the center mass energies around $Z$-boson mass, is
called as a Z-factory, so neutrino production in such a Z-factory
is just the production via $e^-e^+$ collision at $Z$-boson peak.
In fact, since at a $Z$-factory the production of
all kinds of Standard Model (SM) fermion-antifermion pair (except
$t$-quark pair) is mainly via $Z$ boson $s$-channel
annihilation, so it is greatly enhanced by the resonance
effects of the $Z$-boson, i.e. the cross-sections of the
production appear a peak at collision CM energy around $Z$-boson
mass, moreover, the production at leading order is of two
body in final state, so the produced fermion and anti-fermion have definite
momentum and energy. Considering the characteristics,
the production of the three kinds of normal light neutrinos
is expected to be interesting, e.g. the restriction on the
room for the mixing of the light neutrinos with externals, and to
answer whether or with what a luminosity of the Z-factory
the produced neutrinos may be detected with an accessible
detector, even used as a mono-energy
neutrino source etc.

Of the production of the three kinds of normal light neutrinos,
the $\nu_e\bar{\nu}_e$ pair production is special: it is produced
via the interference of the $Z$ boson $s$-channel annihilation and
$W$ boson $t$-channel exchange, although of the rest two species,
the production of $\nu_\mu\bar{\nu}_\mu$ and $\nu_\tau\bar{\nu}_\tau$,
is via the $Z$ boson $s$-channel annihilation as the other fermions
(quarks, muon and $\tau$-lepton). Moreover,
if the invisible width of $Z$-boson measured at
$Z$-factory indeed is due to the contributions from neutrino
production and it may be measured accurately enough,
then not only the number of the light neutrino species
may be realized, but also one may see how big a `room' still left
in the invisible width for the mixing of the normal light neutrinos 
with possible external one(s), whereas the possible mixing is an 
absorbing topic on neutrino physics\cite{sminov,BL1,GN03,MV04}. 
To determine the number of the light neutrino species
and to examine the room left for mixing with externals both is to 
compare the invisible width of $Z$-boson measured at $Z$-factory 
experimentally with the contributions from production of each specie 
of the neutrinos, which are estimated theoretically. Thus we are 
now interested in studying the production of the three species of the 
neutrinos in a $Z$-factory theoretically with care.

The earlier $Z$-factories, such as LEP-I and SLC, run just under a
luminosity of $10^{31}\,{\rm cm^{-2}s^{-1}}$ and
via measuring the `invisible width' of the $Z$ boson, the valuable
conclusion that there are three species of light neutrinos is
obtained. Namely the effective number of light neutrino species:
$N^\nu_{eff}=R^0_{inv}(\frac{\Gamma^Z_{l\bar{l}}}
{\Gamma^Z_{\nu_l\bar{\nu_l}}})_{_{SM}}=2.9840\pm 0.0082$ is obtained
via the experimental measurements of invisible width ratio of $Z$-boson:
$R^0_{inv}\equiv \frac{\Gamma^Z_{inv}}{\Gamma^Z_{l\bar{l}}}=5.943\pm0.016$
at the earlier $Z$-factories\cite{CERN,JB12}. Here if ignoring
the errors, here the effective number of light neutrino species being not an 
integer and smaller than three means the light neutrino species can be 
three but there is some room left for the mixture of the light neutrinos
with external(s). If a new $Z$-factory, called as a super one,
with a luminosity around $10^{35}\sim10^{36}\,{\rm cm^{-1}s^{-1}}$
and proper improved detectors, that is accessible now under the present
technology, is expected to be built, the invisible width of $Z$-boson 
will be measured more accurately (the errors are expected to be suppressed 
greatly), so the conclusion on the room left for light neutrinos mixing with
externals will be improved. Bearing the possible progress in collider and 
detectors in mind, in this paper we will, based on SM, precisely calculate the 
neutrino production, and with the precise and fresh results we will 
discuss (explore) the meaning of refined measurements of the neutrino 
production in a super $Z$-factory with proper detectors in future.

The paper is organized as follows: after INTRODUCTION in section II, we present
formulas of the neutrino production at a Z-factory; in section III we present the numerical results
for the neutrino production; the section IV is to contribute to understanding the results and
discussions.

\section{Neutrino Pair Production at $Z$-factory}

The neutrino production at a $Z$-factory\cite{VN99,PS95}:
\[
    e^-\,e^+\to\nu_i\,\bar{\nu}_i\;\; (i=1,2,3) \,;
\]
where
\[
    \nu_i=\sum_{j=e,\mu,\tau}V_{ij}\;\nu_j\,,
\]
where $V_{ij}$ is the elements of neutrino mixing matrix.
The Standard Model (SM) gives rise to the relevant interaction
between $Z$-boson and leptons:
\begin{eqnarray}
    \mathcal{L}_{\rm int} = gZ_\mu J^\mu_Z,
\end{eqnarray}
where $J^\mu_Z$, the so-called `weak neutral current', is
\begin{eqnarray}
J^\mu_Z&\displaystyle = \frac{1}{2\cos\theta_w}\sum_{i=e,\mu,\tau}
\big[ \bar{\nu}_i\gamma^\mu\frac{1-\gamma^5}{2} \nu_i \nonumber\\
&\displaystyle+\bar{\psi}_i\gamma^\mu (-\frac{1}{2}+2\sin^2\theta_w
    +\frac{\gamma^5}{2})\psi_i\big]\,.\nonumber
\end{eqnarray}
Here another relevant interaction is that between $W^\pm$ and leptons:
\begin{eqnarray}
    \mathcal{L}_{\rm int} = g (W^+_\mu J^\mu_W + W^-_\mu J_W^{\mu\dagger})
\end{eqnarray}
where $J^\mu_W$, the charged weak current, is
\[
    J^\mu_W = \frac{1}{\sqrt{2}} \sum_{i=e,\mu,\tau} \bar{\nu}_i\gamma^\mu
    \frac{1-\gamma^5}{2}\psi_i\,.
\]
At the tree level, there are two relevant Feynman diagrams for the
process $e^-e^+\to\nu_e\bar{\nu}_e$, which are shown in Fig.1.
For the processes,
$e^-e^+\to\nu_l\bar{\nu}_l$ ($l=\mu, \tau$), there is one
Feynman diagram only which is shown in Fig.2.
\begin{figure}
\begin{center}
    \includegraphics[scale=0.6]{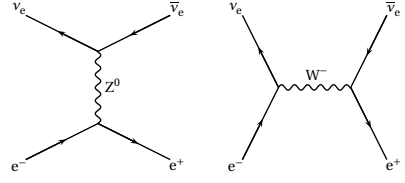}
    \caption{The Feynman diagrams for the process $e^-e^+\to \nu_e \bar{\nu}_e$}. \label{ee}
\end{center}
\end{figure}
\begin{figure}
\begin{center}
\includegraphics[scale=0.6]{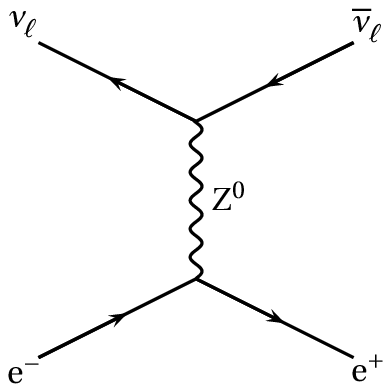}
    \caption{The Feynman diagram for the process $e^-e^+\to \nu_l \bar{\nu}_l$} ($l=\mu, \tau$). \label{mmu}
\end{center}
\end{figure}
It is straightforward to write down the amplitudes according to the
diagram(s). We have
\[
    i\mathcal{M} = \frac{i^2g^2}{2!}({\rm I} + {\rm II})
\]
where
\begin{align*}
    {\rm I}& = -\frac{D^{(Z)}_{\mu\nu}(q)}{2\cos^2\theta_w}\,\bar{u}
    (k)\gamma^\mu\frac{1-\gamma^5}{2}v(k')\,\\
    &\cdot \bar{v}(p')
    \gamma^\mu\left(-\frac{1}{2}+2\sin^2\theta_w+\frac{\gamma^5}
    {2}\right)u(p)\,;
\end{align*}
\[
    {\rm II} = D^{(W)}_{\mu\nu} \bar{u}(k)\gamma^\mu\frac{1-\gamma^5}{2}
        u(p)\,\bar{v}(p')\gamma^\nu\frac{1-\gamma^5}{2}v(k')\,,
\]
and $q = p+p'=k+k'$.
Namely now the whole amplitude for the process $e^-e^+\to\nu_e\bar{\nu}_e$
is shown
\begin{align*}
i\mathcal{M}& = \,\frac{g^2}{2}\Bigg(\frac{D^{(Z)}_{\mu\nu}(q)}{2\cos^2\theta_w}\,
\bar{u}(k)\gamma^\mu\frac{1-\gamma^5}{2}v(k')\,\\
&\bar{v}(p')\gamma^\nu\left(
-\frac{1}{2}+2\sin^2\theta_w+\frac{\gamma^5}{2}\right)u(p)\\
& - D^{(W)}_{\mu\nu}(q')\,\bar{u}(k)\gamma^\mu\frac{1-\gamma^5}{2}u(p)
\,\bar{v}(p')\gamma^\nu\frac{1-\gamma^5}{2}v(k')\Bigg)\,.
\end{align*}
Here to do the calculation, we take the unitary gauge for weak bosons,
so the propagator for $Z$-boson:
\[
D^{(Z)}_{\mu\nu}(q) = \frac{i(-\eta_{\mu\nu}+q_\mu q_\nu/m_Z^2)}{q^2-m_Z^2+i\Gamma_Z m_Z}\,,
\]
and that for $W$-boson:
\[
D^{(W)}_{\mu\nu}(q) = \frac{i(-\eta_{\mu\nu}+q_\mu q_\nu/m_W^2)}{q^2-m_W^2+i\Gamma_W m_W}\,.
\]
Here $\Gamma_Z$, $m_Z$ and $\Gamma_W$, $m_W$ appearing in the denominators are the
total widths, masses of $Z$ and $W$ boson respectively.
Whereas the amplitude for the process $e^-e^+\to\nu_l\bar{\nu}_l\;$ ($l=\mu, \tau$)
is shown
\begin{align*}
i\mathcal{M}& = \,\frac{g^2}{2}\frac{D^{(Z)}_{\mu\nu}(q)}{2\cos^2\theta_w}\,
\bar{u}(k)\gamma^\mu\frac{1-\gamma^5}{2}v(k')\,\\
&\bar{v}(p')\gamma^\nu\left(
-\frac{1}{2}+2\sin^2\theta_w+\frac{\gamma^5}{2}\right)u(p).
\end{align*}

Hence the differential cross-sections
for unpolarized incoming beams are related to the amplitudes via
\[
d\sigma = \frac{1}{2E_1 2E_2 |v_1-v_2|}\overline{\sum}_{spins} |\mathcal{M}|^2\, d\Pi\,,
\]
i.e. in C.M. system for the process $e^-e^+\to \nu_e\bar{\nu}_e$ it is
\begin{eqnarray}\label{differ1}
&\displaystyle\frac{d\sigma}{d\Omega} = \frac{\alpha^2s}{16sin^4\theta_w}\Bigg[
\frac{\cos^4\frac{\theta}{2}}{\left(s\sin^2\frac{\theta}{2}+m_W^2\right)^2+\Gamma_W^2m_W^2}
\nonumber\\
&\displaystyle-(2-\frac{1}{\cos^2\theta_w})\nonumber\\
&\displaystyle\cdot\frac{\left[\Gamma_Z m_Z \Gamma_W m_W-(s-m_Z^2)\left(s\sin^2\frac{\theta}{2}+m_W^2\right)
\right]\cos^4\frac{\theta}{2}}{\left[(s-m_Z^2)^2+\Gamma_Z^2 m_Z^2\right]\cdot\left[\left(s\sin^2\frac{\theta}{2}+m_W^2\right)^2+\Gamma_W^2 m_W^2\right]}\nonumber\\
&\displaystyle+\frac{1}{\cos^4\theta_w}
\frac{\sin^4\theta_w\sin^4\frac{\theta}{2}+(\frac{1}{2}-\sin^2\theta_w)^2\cos^4\frac{\theta}{2}}
{(s-m_Z^2)^2+\Gamma_Z^2m_Z^2}\Bigg]\,;
\end{eqnarray}
for the process $e^-e^+\to\nu_l\bar{\nu}_l$
($l=\mu,\; \tau$) it is
\begin{eqnarray}\label{differ2}
\frac{d\sigma}{d\Omega} = \frac{\alpha^2s}{\sin^4 2\theta_w}
\frac{\sin^4\theta_w\sin^4\frac{\theta}{2}+(\frac{1}{2}-\sin^2\theta_w)^2\cos^4\frac{\theta}{2}}
{(s-m_Z^2)^2+\Gamma_Z^2m_Z^2}\,,
\end{eqnarray}
where
\[
g = \frac{e}{\sin\theta_w}\,,\;\;\alpha =\frac{e^2}{4\pi}
\]
are adopted.
Moreover, from Eq.(\ref{differ1}) one may see that the contributions to the cross-section
from $t$-channel of $W$-scattering and $s$-channel of $Z$-annihilation are destructive.

The total cross-sections can be calculated by integrating
the relevant differential ones respectively.

To calculate out the cross-sections numerically, we take the parameters,
which appear in the formula, from PDG\cite{JB12}:
\begin{eqnarray}
& m_Z=91.1876\, GeV\,,\;\; \Gamma_{Z}=2.4952\, GeV\,,\;\;\nonumber\\
& m_W=80.385\, GeV\,,\;\; \Gamma_W=2.085\, GeV\,,\;\;\nonumber\\
&\sin^2\theta_W=0.23116\; (\,or\, 0.2231)\,,\;\; \alpha=1/127.944\,, \nonumber\\
& m_{\nu_e}\simeq m_{\nu_e}\simeq m_{\nu_e}\simeq 0.0\, eV\,,
\end{eqnarray}
\begin{table}[!h]\label{tab:cross_section}
\renewcommand\arraystretch{1.4}
\centering
\caption
{Total unpolarized cross
 sections vs the c.m. energy $\sqrt{s}$, where $\sigma_1$
 denotes the results with
 $\sin^2\theta_W=0.2312$, while $\sigma_2$ denotes those
 with $\sin^2\theta_W=0.2231$. $\nu_l$ denotes $\nu_\mu$
 and $\nu_\tau$.}
 \begin{tabular}{c|c|c|c|c|c|c}
  \hline\hline
   $\sqrt{s}\;[\mathrm{GeV}]$ &
   89 & 90 & 91 & 92 & 93 & 94 \\ \hline
   $\sigma_1(\nu_e\bar{\nu}_e) \;[\mathrm{nb}]$ &
   0.8101 & 1.8814 & 3.8314 & 3.0146 & 1.4988 & 0.8305 \\ \hline
   $\sigma_1(\nu_l\bar{\nu}_l) \;[\mathrm{nb}]$ &
   0.9477 & 2.0470 & 3.8769 & 2.8379 & 1.3134 & 0.6779 \\ \hline
   $\sigma_2(\nu_e\bar{\nu}_e) \;[\mathrm{nb}]$ &
   0.8514 & 1.9834 & 4.0500 & 3.1943 & 1.5916 & 0.8836 \\ \hline
   $\sigma_2(\nu_l\bar{\nu}_l) \;[\mathrm{nb}]$ &
   1.0023 & 2.1648 & 4.1001 & 3.0013 & 1.3890 & 0.7170 \\ \hline\hline
 \end{tabular}
\end{table}
i.e. the values of the parameters are taken to be renormalization
at $Z$ pole\footnote{Here we do the calculation only at tree level,
thus to have comparatively better results we take the values
renormalized at $Z$-pole for the parameters, but to see uncertainties and
for comparison, when doing the calculation of the total cross-sections
we take a different value of $sin^2\theta_W$, i.e. $sin^2\theta_W=0.23116$ and
$sin^2\theta_W=0.2231$ in different renormalization schemes in cases.
Whereas in the rest calculations of the paper
without declaration, we will take $sin^2\theta_W=0.23116$ only.}.
Precisely the values of the total unpolarized cross-sections
vs C.M.S. energies $\sqrt s$ are collected in TABLE I, and
the curves of the cross-sections are plotted in FIG.3. In order to see the
the difference between the production of $e^-e^+\to\nu_e\bar{\nu}_e$
and that of $e^-e^+\to\nu_l\bar{\nu}_l$ ($l=\mu\,,\tau$), we also plot $\Delta \sigma$,
the difference in total cross-sections, vs $\sqrt s$
in FIG.4.
\begin{figure}\label{E-total}
\begin{center}
\includegraphics[width=2.8in]{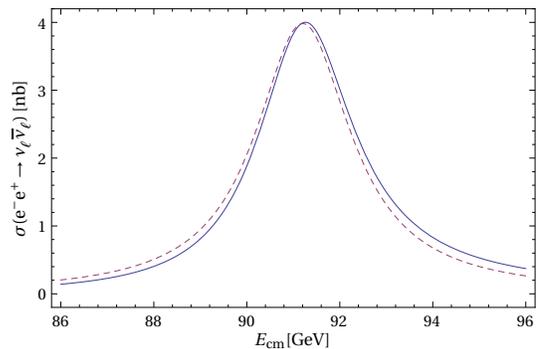}
\caption{The dependence of the total unpolarized cross-sections
on the C.M.S. energies: the
dished curve is those of the processes $e^-e^+\to\nu_l\bar{\nu}_l$
($l=\mu\,,\tau$) and the solid curve is that of the processes
$e^-e^+\to\nu_e\bar{\nu}_e$.}
\end{center}
\end{figure}
\begin{figure}[!htbp]\label{fig:extra-section}
\centering
\includegraphics[scale=0.7]{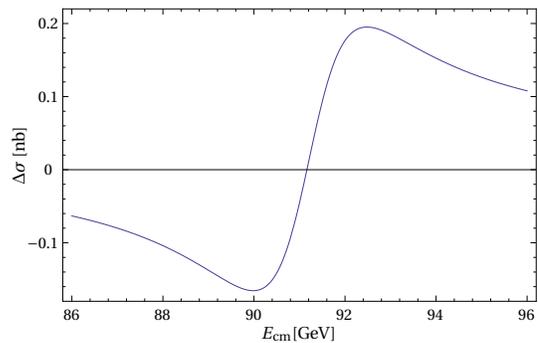}
\caption{The difference $\Delta \sigma$ in the
total cross sections between those of $e^-e^+\to \nu_e\bar{\nu}_e$ and
$e^-e^+\to \nu_l\bar{\nu}_l$ ($l=\mu,\tau$). Here $\Delta \sigma = 0.0$ is
located at$\sqrt{s}\simeq 91.163$GeV ($m_Z$).}
\end{figure}
\begin{figure}[!htbp]
\centering
\includegraphics[scale=0.8]{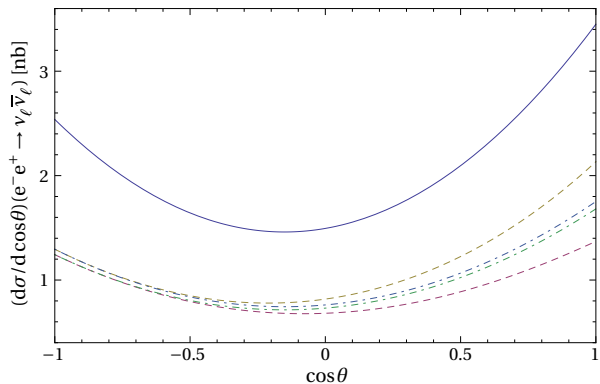}
\vspace{4mm}
\caption{\label{fig:diff-section}The differential cross-sections
for the production $e^-e^+\to\nu_l\bar{\nu}_l$. The solid
curve denotes all the production at $\sqrt s=m_Z$.
The upper and lower dashed curves denote the production for $l=e$ at
$\sqrt s=m_{Z}\pm 0.5\Gamma_{Z}$ respectively.
The upper and lower dot-dashed curves denote the production for $e^-e^+\to
\nu_\mu\bar{\nu}_\mu$ or $e^-e^+\to \nu_\tau\bar{\nu}_\tau$ at
$\sqrt s=m_{Z}\pm 0.5\Gamma_{Z}$ respectively.}
\end{figure}
From the table TABLE I and the figures FIG.3, FIG. 4 we may see that i). The
difference caused by a different value of $\sin^2\theta_W$ is remarkable;
ii). The differences between the total cross-sections for process
$e^-e^+\to \nu_e\bar{\nu}_e$ and for processes $e^-e^+\to \nu_l\bar{\nu}_l$,
$l=\mu\,,\tau$ are tiny at the $Z$-pole, but become sizable
when C.M. energy $\sqrt s$ is away from the $Z$-pole. The differences
are the consequences of the interference of $Z$-annihilation
and $W$-exchange for $e^-e^+\to \nu_e\bar{\nu}_e$ (see FIG.1)
and there is $Z$-annihilation only for $e^-e^+\to \nu_l\bar{\nu}_l$,
$l=\mu\,,\tau$ (see FIG.2);
iii). The interference of the $Z$-annihilation and the $W$-exchange
in the process $e^-e^+\to \nu_e\bar{\nu}_e$) is
`constructive' when $\sqrt s\geq m_Z$, but it is
`destructive' when  $\sqrt s\leq m_Z$ (FIG.4).
\begin{figure}[!htbp]
\centering
\includegraphics[scale=0.7]{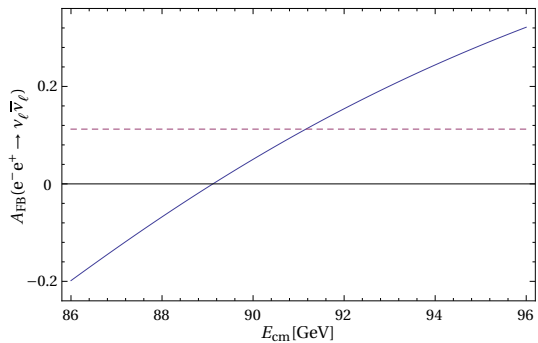}
\caption{\label{fig:FB-asymmetry}The forward-backward asymmetry
$A_{FB}$ for $e^-e^+\to \nu_l\bar{\nu}_{l}$ as a function of $\sqrt s$.
The solid curve presents that of the production of
$e^-e^+\to \nu_e\bar{\nu}_{e}$, and the dashed curve presents that of the production of
$e^-e^+\to \nu_\mu\bar{\nu}_{\mu}$ or $e^-e^+\to \nu_\tau\bar{\nu}_{\tau}$.}
\end{figure}
To see the characters of the production, we have also calculated
the differential cross-sections at three C.M. energies
for the processes $e^-e^+\to \nu_l\bar{\nu}_l$, $l=e\,,\mu\,,\tau$
numerically in C.M.S., and plot them vs the angle $\theta$
between the directions of
the electron beam and the produced neutrinos in FIG.5.
From the figure one may see that the differential
cross-sections are clearly asymmetric in forward and
backward. They are favoured in forward direction and
have a minimum at $\theta\approx\pi/2$.
To highlight the asymmetry, we plot
$A_{FB}\equiv \frac{(\int_{1>cos\theta >0}
-\int_{0> cos\theta >-1}) d\sigma}
{\int_{1>cos \theta >-1}d\sigma}$
vs the C.M. energy $\sqrt s=E_{cm}$
around $Z$-boson mass $m_Z$ in FIG.6.
It is interesting that owing to the interference
of $Z$-annihilation and $W$-exchange the forward-backward asymmetry $A_{FB}$
of the production $e^-e^+\to \nu_e\bar{\nu_e}$ changes sign from minus to
plus as the collision energy increasing, but that of the production
$e^-e^+\to \nu_e\bar{\nu_e}$ keeps constant (see FIG.6).

\section{Conclusions and discussions}
The differential cross-sections for the neutrino
pair production: $e^-e^+\to\nu_e\bar{\nu}_e$, $e^-e^+\to\nu_{\mu}\bar{\nu}_{\mu}$
and $e^-e^+\to\nu_{\tau}\bar{\nu}_{\tau}$ around the CM energy region of
$Z$-boson mass (at a $Z$-factory) are studied thoroughly.
Since there are two Feynman diagrams: $t$-channel exchange and
$s$-channel annihilation
contributing to the first process (FIG.1), but there is only one Feynman diagram,
$s$-channel annihilation contributing to the second and
third ones (FIG.2), thus we have paid more attention on the differences
among the processes.
Since the processes being concerned around $\sqrt S\simeq
m_Z$, the $t$-channel exchange diagram's contribution is much smaller than
that from the $s$-channel annihilation. The differences between the
production $e^-e^+\to\nu_e\bar{\nu}_e$ and $e^-e^+\to\nu_{\mu}\bar{\nu}_{\mu}$
or $e^-e^+\to\nu_{\tau}\bar{\nu}_{\tau}$ are tiny at $Z$-boson peak, but
become visible when off the peak (FIGs.3-5 and TABLE.I). Namely the shape
of the $Z$-boson resonance for the production process $e^-e^+\to\nu_e
\bar{\nu}_e$ around the $Z$-boson peak is distorted by
the two diagram interference in certain degree. Therefore
we think that the facts described here should be treated very carefully,
especially when someone considers the room left for the mixing of the
three light neutrinos with heavy neutral leptons and/or stereos via
taking the invisible width of $Z$-boson into account. Precisely we mean that
if one would like to obtain the effective number $N^\nu_{eff}$ of the light neutrino
species and to suppress the relevant errors, the way to obtain it via the
data of the earlier $Z$-factories\cite{CERN,JB12} below:
$$N^\nu_{eff}=R^0_{inv}(\frac{\Gamma^Z_{l\bar{l}}}
{\Gamma^Z_{\nu_l\bar{\nu_l}}})_{_{SM}}=2.9840\pm 0.0082\,,$$
$$R^0_{inv}\equiv \frac{\Gamma^Z_{inv}}{\Gamma^Z_{l\bar{l}}}=5.943\pm0.016$$
should be added the distortion effects being affected carefully.

Since the neutrino-antineutrino pair production by electron-positron
annihilation at $Z$-boson resonance is of a two-to-two body process
and with resonance enhancement, so the produced neutrino and antineutrino
at a $Z$-factory are productive and of mono-energy. Moreover $Z$-boson mass
is quite heavy $m_Z\simeq 91.2$GeV, so roughly the energy of the produced neutrino
$E_\nu=\frac{m_Z}{2}\simeq 45.6$GeV (tens GeV order).
Thus the produced neutrinos with the character: a quite high energy and mono-energy,
seems may find some special usages in principle.

Based on the estimates of the production
cross-sections in this paper (TABLE.I) and the differential
cross-sections (FIG.5) quantitatively, one may realize
that indeed the cross-section is greater
in the forward direction than those in the other directions,
thus if the produced neutrinos are considered as beams in directions
then the beam intensity in forward direction is biggest.
Namely, considering the cross-sections of neutrino with matters
are very tiny, if one would like to detect the produced neutrinos
directly and successfully, then the most hopeful way is
to put the detector at the forward direction. Moreover,
the production cross-sections at a $Z$-factory
are in the order of a few $nb$ as shown in TABLE.I and FIG.5, therefore
only when the luminosity of the $Z$-factory is higher than
$10^{36}$ cm$^{-2}$s$^{-1}$, so that the flux of the produced neutrinos
may be great enough for an accessible detector\footnote{The cross-sections
of the energetic neutrinos
(several tens GeV in energy) colliding with common matter are so small in the
magnitude order of $10^{-36}\sim 10^{-37}$cm$^2$, so even with a huge detector,
e.g. a detector with $10$km long (thickness), the neutrinos produced at a $Z$-factory
can be detected only when the luminosity of the $Z$-factory is so high as
pointed here, i.e., the intensity of the produced neutrinos reaches
to the ability of the accessible detector to detect them.}. But considering
the possible progress in technology on $Z$-factory and the detectors, one
may dream to apply the mono-energy neutrinos produced at
a $Z$-factory to doing experiments on neutrino physics one day in future.

{\bf Acknowledgement}
\indent\indent The author (C.-H. Chang) would like to thank
Prof. Zuo-Xiu He for
valuable discussions on the subject. The work is supported by the
National Natural
Science Foundation of China (NNSFC) with Grant No. 11275243,
No. 11147001, No. 11275036, No. 11047002, No. 11205227;
the open project of State Key Laboratory of Theoretical Physics
with Grant No. Y3KF311CJ1; the Natural Science Foundation of
Hebei province with Grant No. A2013201277; and Natural Science Foundation of
Hebei University with Grant No. 2011JQ05, No. 2012-242.1го

\end{document}